# The effectiveness of altruistic lobbying: A model study

P.Yu. Chebotarev, Z.M. Lezina, A.K. Loginov, Ya.Yu. Tsodikova

Institute of Control Sciences of the Russian Academy of Sciences, Moscow, Russia

## 1. Introduction

Altruistic lobbying is lobbying in the public interest or in the interest of the least protected part of the society. This "or" is the point: it is necessary to bear in mind that an altruist has a wide range of strategies, from behaving in the interest of the society as a whole to the support of the most disadvantaged ones. How can we compare the effectiveness of such strategies? The second question is: "Given a strategy, is it possible to assess the optimal number of participants choosing it?" Finally, do the answers to these questions depend on the level of well-being in the society? For example, can we say that the poorer the society, the more important is to focus on the support of the poorest? We answer these questions within the framework of the model of social dynamics determined by voting in a stochastic environment [1–5].

Consider a society consisting of $n$ members (agents); each of them is characterized by a real scalar interpreted as the utility (or capital) level; a negative value is naturally interpreted as a debt. Let the initial distribution of capital be given. A *proposal of the environment* is a vector of increments of individual capitals. Let these increments be random variables; in the simplest case, we assume that they are independent and identically distributed with mean $\mu$ and standard deviation $\sigma$. The environment generates a series of proposals each of which is put to the vote. Every agent takes part in the voting and has one vote. With some voting procedure, the profile of votes is converted into a collective decision: the proposal is approved or rejected. The approved proposals are implemented: the participants get the capital increments specified in the proposal. Considering a large series of votes, one can explore the dynamics of the vector of agents' capital: in different environments, with different social attitudes, and with various voting procedures. An interesting version of the model is that with ruining: the agents whose capital values become negative are ruined.

## 2. A study of altruistic lobbying

Consider a society consisting of two categories of agents: selfish and altruistic ones. An egoist votes for a proposal if and only if this proposal increases his capital.

As noted in the introduction, the altruists have a wide range of strategies. Consider the following type of strategy. We order all agents by the increase of the current capital value. Since the capital increments are real-valued random variables, we can assume that all capitals are different. Let an integer $n_0 \leq n$ be fixed. For the current proposal, one calculates the total capital increment of $n_0$ poorest agents. An altruist supports the proposal if and only if this total is positive, no matter what happens to his own capital. If $n_0 = n$, then the altruist supports those proposals that enrich the society as a whole. When $n_0$ is smaller, then he supports more or less numerous "lower" stratum of the society.

Let $K$ be the same initial capital of all the agents. Assume that the distribution of capital increments is Gaussian: $N(\mu, \sigma)$. Consider the case where bankrupt agents drop out. Let $m$ be the number of steps of voting constituting the "game". The strategy of altruists is determined by $n_0$. As a criterion of the effectiveness of this strategy we consider the relative number of participants that "survive" till the end of the game. Another important parameter is the number (percentage) of altruists in the society.

We study the following question: "How the effectiveness of altruists' strategy depends on their proportion and the threshold $n_0$ at different parameters $\mu$ and $\sigma$?" In other words, "What is the optimal proportion of altruists in the society and what is their optimal strategy in a favorable ($\mu > 0$), neutral ($\mu = 0$), and unfavorable ($\mu < 0$) environment?"

Some simulation results are as follows. Let the society consist of $n = 100$ agents; the number of altruists, $m$, varies. The experiments are conducted at $\sigma = 12$; game lasts 500 steps. On the horizontal axes in Fig. 1, we have the number of altruists, $m$, and their strategy parameter $n_0$; along vertical axis, the relative number of agents at the end of the game. In the diagrams, $\mu$ and the initial capital $K$ of the agents are also shown. Some results are as follows.

1. In a neutral environment ($\mu = 0$) and with the support of the poorest ($n_0$ is small), the number of altruists should be small, or they waste their influence, resulting in ruining players outside the "support screen". At a high $n_0$, we have saturation with the increase of the number of altruists. With $K = 40$, the optimum is reached at $n_0 = 50$ and $m > 40$ (at higher values of $m$, saturation holds). In other words, altruists should support the lower half of the community. With the extension of the support screen, the effectiveness of the altruistic strategy declines.

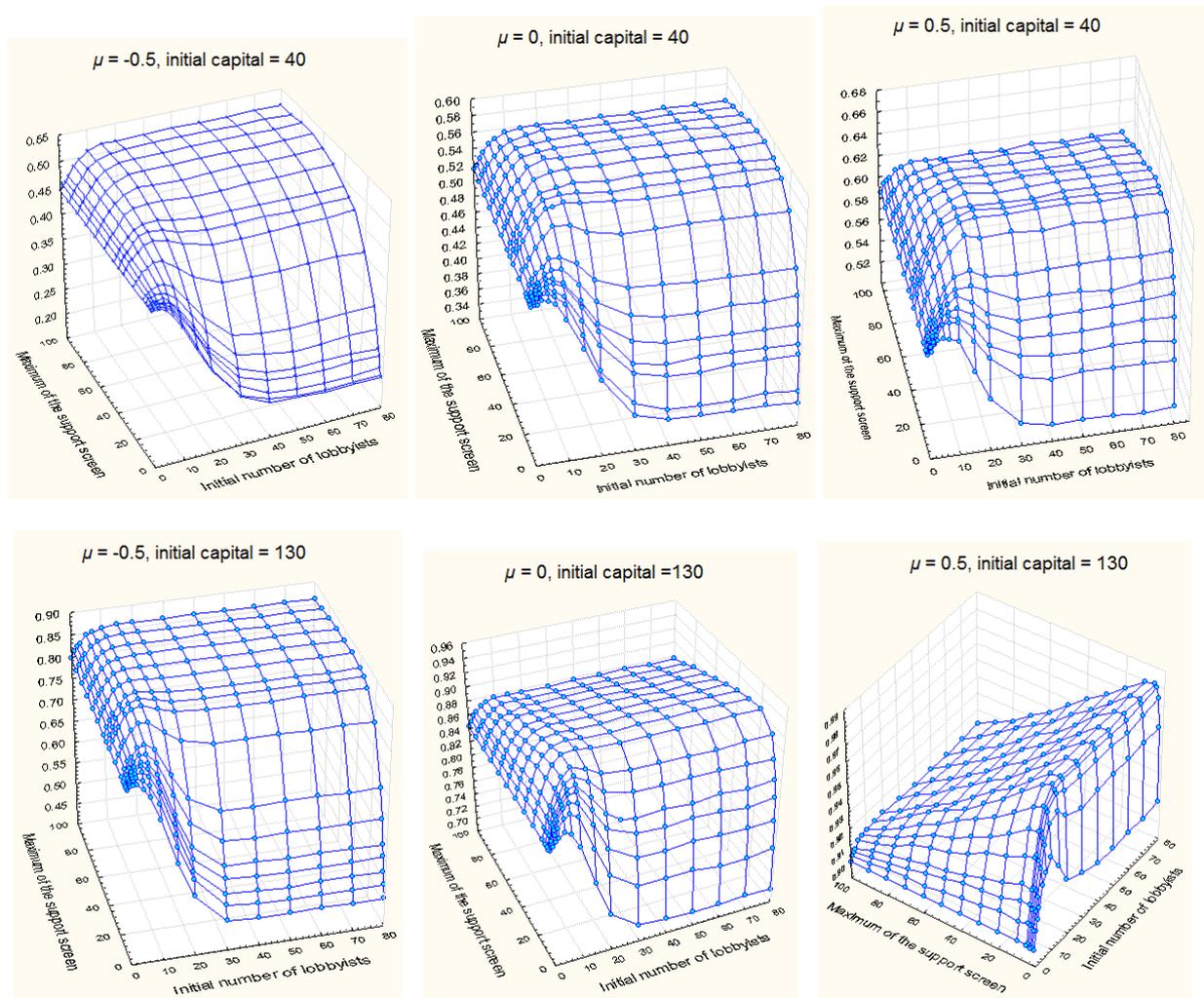

Fig. 1.

2. In an unfavorable environment ($\mu = -0.5$) and a high initial capital ($K = 130$), the picture is very similar. The only significant difference is that there is no pronounced effect of reducing the effectiveness of the strategy with the increase of the right border of the support screen. Maximum is reached at $n_0 = 30$ and $m = 10$, but it does not change significantly with an increase of these parameters.
3. In an unfavorable environment ($\mu = -0.5$) and a lower initial capital ($K = 40$), the maximum is reached at $n_0 = 80$ and $m = 60$. This is a very interesting conclusion: in a really harsh environment, we have to support everyone!
4. Finally, in a supportive environment with a high initial capital ($\mu = 0.5$, $K = 130$), the picture changes dramatically: we should support a thin layer of the poorest; the others will "resurface themselves".

The main conclusion is simple: in rich and affluent societies, support of the poorest is effective. The poorer is the society, the more "wasteful" is the support of the poorest. If you focus on this support, then the zone of trouble grows. A more detailed analysis of the results reveals some more subtle phenomena as well.